\documentclass[10pt,letterpaper,twocolumn,aps,showpacs,preprintnumbers,amsmath,amssymb,nofootinbib,prl]{revtex4-2}
\usepackage{graphicx}
\usepackage{dcolumn}
\usepackage{bm}
\usepackage[utf8]{inputenc}
\usepackage[T1]{fontenc}

\usepackage{color}
\usepackage{amsmath}
\usepackage{amssymb}
\usepackage[unicode=true,pdfusetitle, bookmarks=true,bookmarksnumbered=false,bookmarksopen=false,
 breaklinks=false,pdfborder={0 0 0},pdfborderstyle={},backref=false,colorlinks=true]
 {hyperref}
\hypersetup{
 pdfborderstyle={},colorlinks,linkcolor=red,citecolor=blue}
\makeatletter

\usepackage{setspace}
\usepackage{mathrsfs}
\usepackage{siunitx}
\usepackage{soul}

\makeatother
\begin{document}
\title[\texorpdfstring{$\quad$}{}]{Generation of a CW anti-bunched photon field from a thin-film PPLN waveguide by two-photon interference with a weak coherent state}

\author{Yue Li\texorpdfstring{$^{1,2,*}$}{}, Haochuan Li\texorpdfstring{$^{1,*}$}{}, Yuhang Lei\texorpdfstring{$^{2}$}{},  Xiaoting Li\texorpdfstring{$^{1}$}{}, Jianmin Wang\texorpdfstring{$^{2}$}{}, Xuan Tang\texorpdfstring{$^{2}$}{}, \\
Mu Ku Chen\texorpdfstring{$^{1}$}{},
E. Y. B. Pun\texorpdfstring{$^{1}$}{}, Cheng Wang\texorpdfstring{$^{1,\$}$}{}, and Z. Y. Ou\texorpdfstring{$^{2,\dag}$}{}} 

\affiliation{\mbox{\texorpdfstring{$^{1}$}{}Department of Electrical Engineering, City University of Hong Kong, 83 Tat Chee Avenue, Kowloon, Hong Kong}\\
\mbox{\texorpdfstring{$^{2}$}{}Department of Physics, City University of Hong Kong, 83 Tat Chee Avenue, Kowloon, Hong Kong}\\ 
$^*$equal contribution, \texorpdfstring{$^{\$}$}{}cwang257@cityu.edu.hk, \texorpdfstring{$^{\dag}$}{}jeffou@cityu.edu.hk}

\date{\today}

\begin{abstract}
An anti-bunched photon field is produced from a thin-film ppln waveguide by mixing the on-chip two-photon state with a weak but matched coherent state. This is achieved by taking out the two-photon part of the coherent state via a destructive two-photon interference with the on-chip two-photon state. We achieve a photon rate of 100 kHz with a $g_2$-value of 0.35. This anti-bunched light field will have applications in high-resolution quantum imaging such as long baseline quantum telescopy for enhancing the signal-to-noise ratio.
\end{abstract}

\maketitle

Photon anti-bunching effect was the first evidence of non-classical behavior of light fields, showing the particle nature \cite{kimble1977photon}.  It is usually cast as $g_2 < 1$ with $g_2 \equiv \langle{I^2}\rangle/{\langle{I}\rangle}^2$ being the normalized intensity autocorrelation. Anti-bunched light fields were produced mostly from the fluorescence of a single emitter such as a single atom \cite{kimble1977photon}, a trapped single ion \cite{diedrich1987nonclassical,schubert1992photon}, a single molecule \cite{basche1992photon}, and a quantum dot \cite{michler2000quantum}. This is because a single emitter cannot give another photon right after emitting one photon, which leads to the ideal single-photon case of $g_2=0$.

Single photons are good quantum information carriers as the flying qubit.  They were ideal for quantum cryptography due to the quantum no-cloning theorem. High-quality single photons can be used for optical quantum computing and simulation \cite{knill2001scheme,wang2017high}. Thus, persistent engineering efforts have been focused on producing single photons with good temporal and spatial modes \cite{ding2016demand}.

Current state-of-the-art single-photon sources, leveraging techniques such as microcavity-enhanced Purcell effects and resonant excitation, can achieve emission rates at the MHz level. However, this performance critically depends on precise cavity design, spectral stability engineering, and suppression of multi-photon events. Further challenges include maintaining high photon indistinguishability and collection efficiency under these high-rate conditions, which often require cryogenic temperatures and advanced noise-mitigation strategies for practical scalability\cite{ding2025high,tomm2021bright}.

Recently, it was proposed \cite{gottesman2012longer} and tested in a tabletop setting \cite{brown2023interferometric,thekkadath2023intensity} that entangled single photons can be used in quantum telescopy via a two-photon interference effect for high-resolution optical imaging, such as long-baseline astronomy applications. Although coherent states can replace the entangled single-photon state, anti-bunched light such as the single-photon state can significantly enhance the signal rate, which is especially desirable in astronomical applications where the light level is extremely low \cite{tang2025phase}. However, the single-photon sources developed so far are not suitable for astronomical applications because they are mostly in the pulsed mode, while celestial light is in continuous wave (cw) form, leading to temporal mis-match and significantly reduced two-photon interference effect between the celestial photons and single-photon sources. In addition, the fixed emission wavelength of single emitters severely limits their astronomical applications.

 On the other hand, quantum technology of two-photon interference provides another approach for producing cw anti-bunched optical field. It was proposed in 1974 \cite{stoler1974photon} and demonstrated with pulsed fields more than 30 years ago \cite{koashi1993photon,lu2001observation} that mixing a two-photon state with a weak coherent state can cancel the two-photon part of the coherent state, leading to an anti-bunched light field. However, because the single-photon state so produced is still probabilistic and narrow band two-photon states are required, which usually involves a complicated cavity system \cite{ou1999cavity}, the technique is not popular and does not have much use in quantum information sciences. Nevertheless, recent quantum imaging/telescopy application of the single-photon state for sensitivity enhancement does not require the on-demand property \cite{tang2025phase}, so the technique of two-photon interference for single-photon production is more favorable because of its high production rate and good mode profile, its continuous wave nature to match celestial light, and flexibility in wavelength selection through nonlinear optics.

     \begin{figure*}
        \includegraphics[width=17cm]{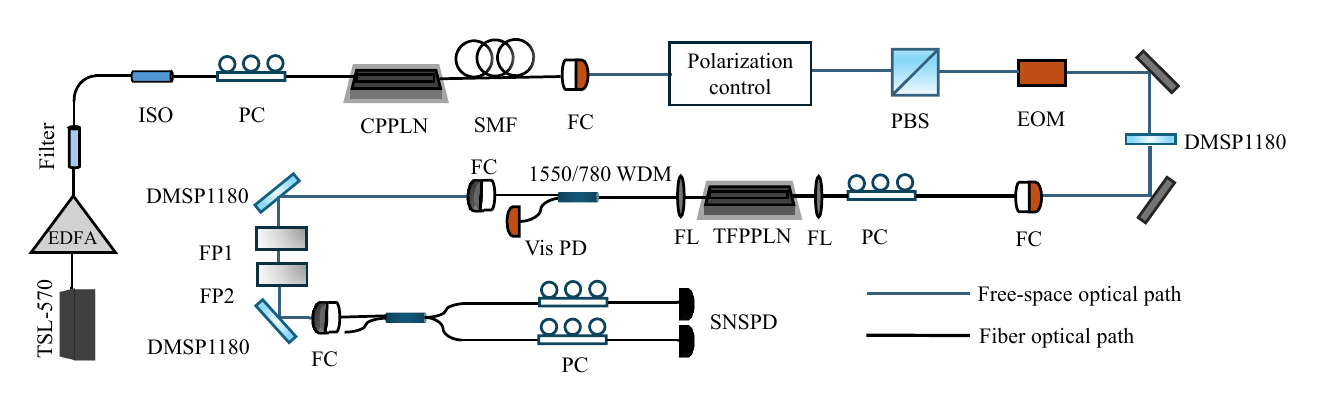}
            \caption{Experimental layout.  EDFA—erbium doped fiber amplifier, CPPLN—commercial PPLN, TFPPLN—thin film PPLN, SMF—single-mode fiber of 780 nm, ISO—isolator, PBS—polarizing beam splitter of 780 nm, EOM—electro-optic phase modulator, DMSP—shortpass dichroic mirror, FC—fiber coupler, PC—polarization control, FL—fiber lens. }
            \label{setup}
    \end{figure*}

In practical implementation, the narrow band requirement of the two-photon state is the main obstacle in the application of the approach because the traditional two-photon source of spontaneous parametric down-conversion is broadband and passive filtering significantly reduces the pair rate so that a complicated active resonator system has to be employed for rate enhancement \cite{ou1999cavity}. However, with the emergence and maturity of waveguide nonlinear optics technology, the photon pair production efficiency has increased dramatically, so we can reconsider the simple method of passive filtering for bandwidth reduction. In this paper, we employ a periodically poled lithium niobate thin-film waveguide with an SHG conversion efficiency of 455{\%}/W  to produce a two-photon state\cite{wang2018ultrahigh}. The high two-photon production rate allows us to employ passive optical filters to reduce the bandwidth to the \SI{200}{\MHz} level. This enables us to implement a time-resolved two-photon destructive interference technique for removal of the two-photon events from a coherent state, thus achieving an anti-bunched optical field at a rate of \SI{100}{\kHz}.

Theory -- It was first suggested by Stoler that a squeezed coherent state can exhibit anti-bunching behavior of light when proper parameters are set \cite{stoler1974photon}. More detailed analysis \cite{lu2005security} shows that it is a two-photon interference effect that cancels the two-photon events, which can originate from either a two-photon state generated by a spontaneous parametric process or the two-photon part of a coherent state. The single-photon part of the coherent state is not affected because the spontaneous parametric process does not produce single-photon events that rival the coherent state.  This can be seen more clearly from the expression of the squeezed coherent state in photon number basis:
\begin{equation}\label{sq-co}
    |\alpha, \eta\rangle = \hat S(\eta) |\alpha\rangle = \sum_{m=0}^{\infty} C_m |m\rangle
\end{equation}
where $\hat S(\eta) = \exp[(\eta^* \hat a^2 - \eta \hat a^{\dagger 2})/2]$ is the squeezing operator and $\hat S(\eta) \approx 1 + (\eta^* \hat a^2 - \eta \hat a^{\dagger 2})/2$ for $\eta\ll 1$. $|\alpha\rangle$ is the coherent state and $|\alpha\rangle\approx 1+\alpha |1\rangle + (\alpha^2/\sqrt{2})|2\rangle$ for $\alpha\ll 1$. Using the approximate forms of $\hat S$, $|\alpha\rangle$ under $\eta, \alpha \ll 1$ and making an expansion in eq.(\ref{sq-co}), we obtain
\begin{eqnarray}
    && C_0 \approx 1, C_1\approx \alpha,~~C_2 \approx (\alpha^2-\eta)/\sqrt{2}, \cr
    && C_3 \approx \alpha(\alpha^2-3\eta)/\sqrt{6}, ....
\end{eqnarray}
Notice that $C_2 = 0$ for $\alpha^2=\eta$, leading to the disappearance of the two-photon term in Eq.(\ref{sq-co}). It is straightforward to find that $g_2\equiv \langle :\hat N^2:\rangle/\langle \hat N\rangle^2 \approx 4|\alpha|^2 \ll 1 (\hat N \equiv \hat a^{\dagger}\hat a)$, leading to anti-bunching effect.

The squeezed coherent state can be produced by injecting a coherent state into a parametric amplifier. In the experimental implementation, we employ a \SI{6}{\mm}-long periodically poled lithium niobate (PPLN) waveguide on a thin-film platform as the $\chi^{(2)}$-nonlinear medium for the parametric process. The manufacture of the thin-film ppln waveguide is based on a mature fabrication process \cite{wang2018ultrahigh,li2024advancing}, where a 600-nm-thick x-cut MgO-doped LNOI wafer is employed with 220nm-etching depth, 60° sidewall angle, and an 800-nm silica cladding. The PPLN waveguide is pumped by a field at 780 nm to produce a pair of down-converted photons at 1560 nm, which is also the wavelength of the injected coherent state. Notice that condition $\alpha^2=\eta$ requires the phase lock between the injected coherent state and the pump field of the two-photon generation whose field amplitude is included in $\eta$. For this, we generate the pump field at 780 nm by using second-harmonic generation of a laser operating at 1560 nm, which also serves as the injected coherent state. The experimental setup is shown in Fig.\ref{setup} where the 1560 nm field from a laser (Santec) is first frequency-doubled with a commercial PPLN waveguide (CPPLN, HC Photonics) to 760 nm. The leftover 1560 nm field is first filtered by a 780 nm single-mode fiber, then passes through  780 nm fiber couplers, and is further attenuated to the few-photon level. The intensity of the 1560 nm field is fine-tuned and controlled by an ensemble of polarization control elements without much reduction of the 780 nm field. The combined field is coupled to and out of the thin-film ppln waveguide by fiber lenses (FL).  The output field passes through a WDM to filter out the 780 nm field. The bandwidth of the filtered field is further narrowed to 200 MHz via two Fabre-Perot (FP) plates with different free spectral ranges before split and sent to two superconducting nanowire single-photon detectors (SNSPD). The digital pulses from SNSPDs are sent to a time tagger (Swabian) for time-resolved coincidence measurement.

    \begin{figure}
       
        \includegraphics[width=8.5cm]{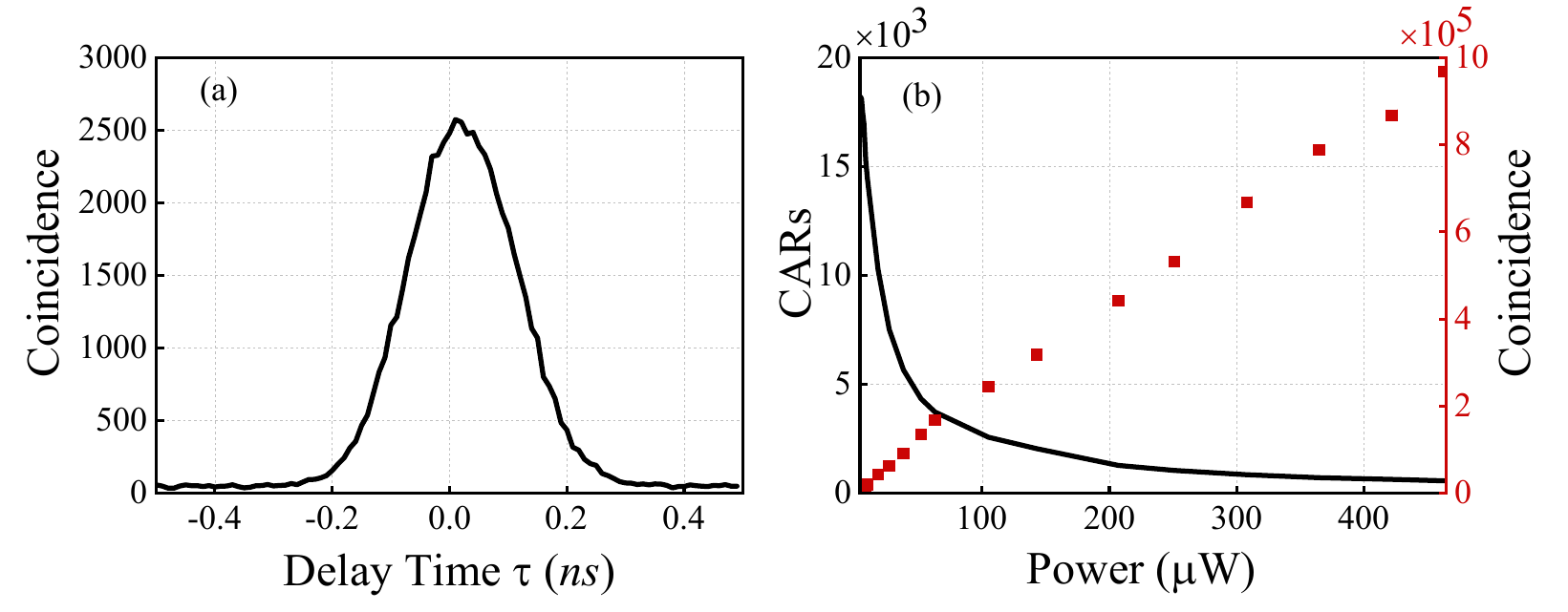}
            \caption{Characterization of the broadband two-photon state directly from the thin-film PPLN source. (a) Time-resolved coincidence in 2 min. (b) Black trace: Coincidence-to-accidental ratio (CAR) and red points: total coincidence in 2 min as a function of pump power.}
            \label{CAR}
    \end{figure}

We first characterize the two-photon state from the thin-film ppln waveguide and test its efficiency by blocking out the 1560 nm field completely with a dichroic mirror (DMSP1180) before coupling on to the thin-film ppln. We start with the broadband two-photon state directly from the TFPPLN without the two narrowband Fabre-Perot filters. The results are shown in Fig.\ref{CAR}, where in (a) we display the time-resolved coincidence measurement.  The coincidence time bin in Fig.\ref{CAR}(a) is 30 ps and the full width at half height is 250 ps, mainly resulting from SNSPD's time jitter uncertainty. The ratio of the peak value to the flat background value on the two sides gives the coincidence-to-accidental ratio (CAR) and is plotted as a function of pump power in Fig.\ref{CAR}(b) as the black trace. The overall coincidence count in 2 min (red) is displayed as well. 

    \begin{figure}
        \includegraphics[width=8cm]{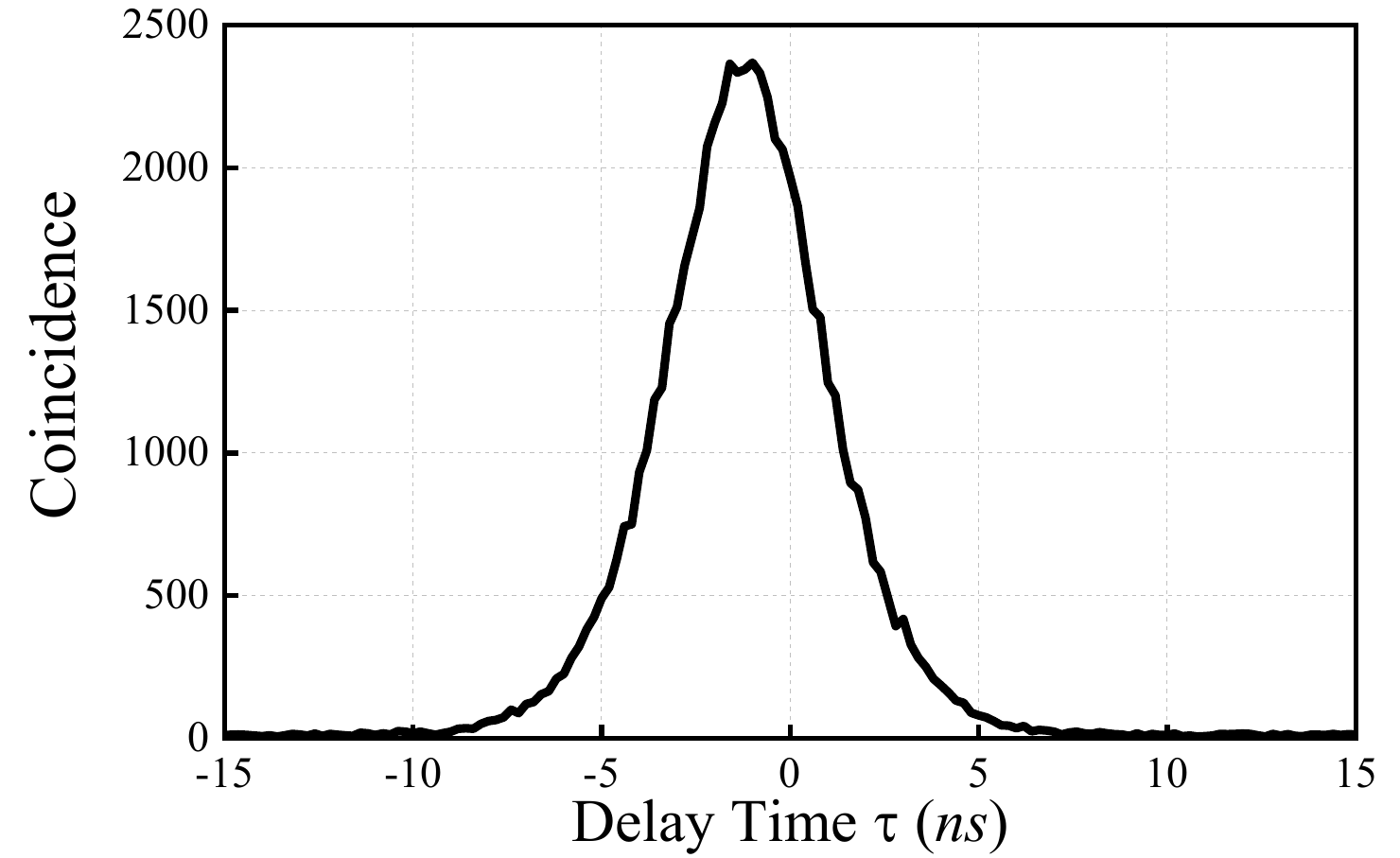}
            \caption{Time-resolved coincidence counts in 50 min or intensity correlation function $\Gamma_2(\tau)$ for the narrow band two-photon state after two Fabre-Perot filters. Time bin is 30 ps.}
            \label{g2}
    \end{figure}

Next, we place the two F-P filters to narrow the bandwidth of the down-converted photon pairs to 200 MHz, leading to a correlation time of 5 ns so the detectors can resolve in time. The measured time-resolved coincidence measurement ($\Gamma_2(\tau)\equiv \langle :\hat I(t)\hat I(t+\tau):\rangle$) is shown in Fig.\ref{g2} where the full width at half height is 5 ns, consistent with the bandwidth of 200 MHz. The time bin is again 30 ps and the total time of data taking is 50 min. Integrating over the whole range gives a total coincidence of 62000, resulting in a detected pair rate of $R_{2ph}=62000/(50\times 60s) = 21 s^{-1}$.

\begin{figure}
        \includegraphics[width=8cm]{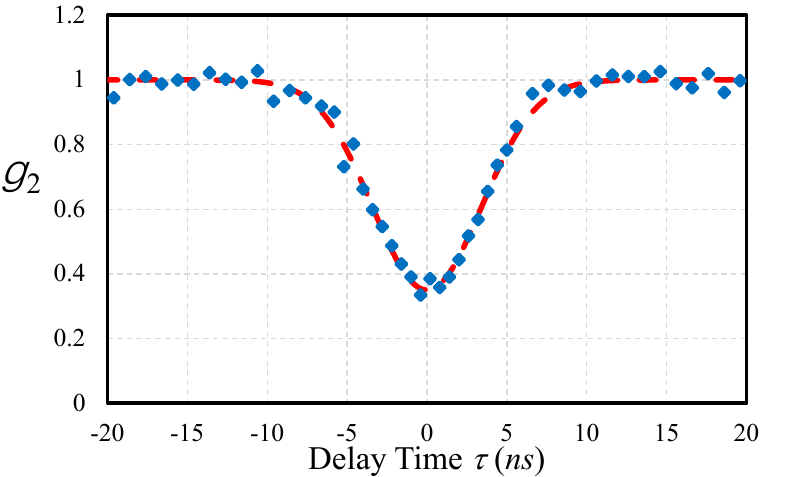}
            \caption{Normalized intensity correlation function $g_2(\tau)$ for the anti-bunched light field generated by mixing the narrow band two-photon state with a weak coherent state. The error bars are the size of the markers. The dashed red line is a fit to Gaussian with a width of $\sigma=3.4 ns$ and a visibility of 0.65. }
            \label{g2-anti}
\end{figure}

The narrow band two-photon state is ready to be mixed with the coherent state injection at the input. For this, we take out the DMSP1180 to allow the 1560 nm coherent field to pass and input into TFPPLN. The relative phase between the pump field and the coherent injection is controlled and fine-tuned by an electro-optic phase modulator (EOM). When the relative phase is set correctly and the intensity of the injected coherent state is properly adjusted to match its two-photon coincidence rate to that of the narrow band two-photon state, we record the time-resolved normalized intensity correlation function $g_2(\tau) \equiv \Gamma_2(\tau)/\Gamma_2(\infty)$. The result is shown in Fig.\ref{g2-anti}, which clearly displays the anti-bunching effect of $g_2(0) < g_2(\tau)$ with a minimum value of $g_2(0) = 0.35$. The red dashed line is a best-fit Gaussian with a width of $\sigma=3.4 \text{ ns}$. The coincidence counts are collected within a bin width of 200 ps  in the central region, and the large fluctuation arises from the relatively low coincidence rate at $\Gamma_2(\infty) \sim 1200/$18 min per data point. The minimum value of $g_2(\tau)$ is $g_2(0) = 0.35$, which is relatively large compared to the theoretical value of 0.0003 (see estimation later), and is believed to be due to phase drift/instability during the long data collection period of more than 1000 s for building up good statistics. In the meantime, we record a single-photon rate of 100 kHz, which is mainly from the contribution of the injected coherent state as indicated from Eq.(\ref{sq-co}). This rate is significantly higher than that of the current single-photon source from single emitters such as ions and quantum dots and can be improved depending on the two-photon rate of SPDC, as shown below.

From the theory presented earlier, we find that the rate of the anti-bunched field ($\propto |\alpha|^2$) is limited by the pair rate ($\propto |\eta|^2$) of the two-photon state: $|\alpha|^2=|\eta|$. Notice that since we use a single-mode model in the theory part,  $|\alpha|^2$ is the average photon number in one mode, which has a temporal size of coherence time $T_c$ for the cw field. So, the photon rate is on the order of $R_{1ph} \sim |\alpha|^2/T_c$. $|\eta|^2$ is the average number of pairs per mode so the pair rate is $R_{2ph} \sim |\eta|^2/T_c$. The condition of $|\alpha|^2=|\eta|$ leads to $R_{1ph} \sim \sqrt{R_{2ph}/T_c}$. In our case here, from Fig.\ref{g2}, $T_c$ is the width of the $\Gamma_2(\tau)$-function, which is 5 ns, and $R_{2ph} = 21 s^{-1}$ is the total two-photon coincidence rate. So, $R_{1ph} \sim \sqrt{R_{2ph}/T_c} \sim 6.4\times 10^4$ Hz, which is consistent with the measured single rate of 100 kHz. This rate can certainly be improved considering the large overall chip coupling loss of more than 10 dB.
However, as the single-photon rate increases, the minimum value of $g_2 =|\alpha|^2 (\sim R_{1ph}T_c \sim \sqrt{R_{2ph}T_c})$ also increases, according to the theory presented earlier, due to higher-order contributions.  In our case here, the theoretical estimate is $g_2 \sim \sqrt{R_{2ph}T_c} = \sqrt{5\times 10^{-9} \times 21} = 3 \times 10^{-4}$, which is very small even if we increase the single rate by 10 folds. Notice that our experimentally observed value of 0.35 is significantly higher than this theoretical value, indicating some technical imperfections, mainly due to phase instability, as we discussed earlier.

\begin{figure}
        \includegraphics[width=8cm]{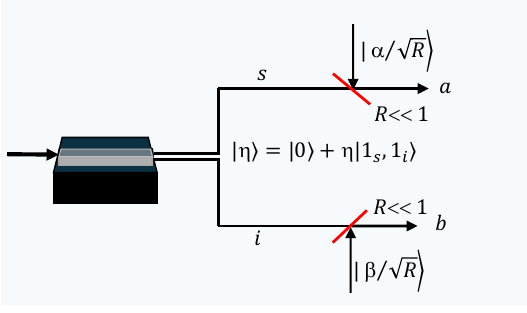}
            \caption{Schematic for generating high rate path-entangled single-photon state.}
            \label{path-ent}
    \end{figure}

Once a single-photon state is produced, a path-entangled single-photon state can be generated with a 50:50 beam splitter with half the rate to each side. There is a variant of the current two-photon interference scheme for generating a single-photon state with path entanglement. Instead of mixing the two-photon state with one coherent state, we can separate the two photons and mix each with individual coherent states, respectively, as shown in Fig.\ref{path-ent}. This scheme was originally used for the demonstration of violation of Bell's inequality \cite{kuz00}. But here by canceling the two-photon state of $|1_a,1_b\rangle$, we can realize a path-entangled single-photon state of $|1_{ab}\rangle = (|1_a, 0_b\rangle + |0_a, 1_b\rangle)/\sqrt{2}$. This can be seen in the following derivation. 

Referring to Fig.\ref{path-ent}, we write, up to two-photon terms, the two coherent states and the two-photon state from the spontaneous parametric process after the highly transmissive beam splitters ($R\ll 1$) as
\begin{eqnarray}\label{state}
|\eta\rangle_{si} &\approx & |0\rangle + \eta |1_s,1_i\rangle \rightarrow |0\rangle + \eta |1_a,1_b\rangle, \cr
|\alpha\rangle_a & \approx & |0_a\rangle + \alpha |1_a\rangle + (\alpha^2/\sqrt{2})|2_a\rangle\cr
|\beta\rangle_b & \approx & |0_b\rangle + \beta|1_b\rangle + (\beta^2/\sqrt{2})|2_b\rangle .
\end{eqnarray}
Then the state of mode $a,b$ becomes
\begin{eqnarray}
|\Phi\rangle_{ab} &=& |\eta\rangle_{si}|\alpha\rangle_a|\beta\rangle_b\cr
&\approx & |0\rangle + \alpha |1_a, 0_b\rangle + \beta |0_a, 1_b\rangle + (\eta + \alpha\beta)|1_a,1_b\rangle \cr
&&~+ (\alpha^2/\sqrt{2}) |2_a, 0_b\rangle + (\beta^2/\sqrt{2}) |0_a, 2_b\rangle\cr
&=& |0\rangle + \alpha (|1_a, 0_b\rangle +|0_a, 1_b\rangle) + \alpha^2 (|2_a, 0_b\rangle \cr
&&~+  |0_a, 2_b\rangle)/\sqrt{2} ~~~~~~~~ {\rm if} ~\beta =\alpha, \eta=-\alpha^2.
\end{eqnarray}
Here, since the transmissivity $T \sim 1$, we ignore the case when photons go to the other side of the beam splitters. Furthermore, we can neglect the two-photon state in the expression above if the coincidence measurement is performed between $a$ and $b$. Then the dominant term is the path-entangled two-photon state $(|1_a, 0_b\rangle +|0_a, 1_b\rangle)$, which can be used for enhancing the signal rate in quantum telescopy applications \cite{gottesman2012longer}.

\vskip 0.2in

In summary, we use the technique of quantum two-photon interference to generate an anti-bunched light field, which is advantageous to the the traditional single-photon sources from single emitters in that it has a high photon flux with a well-defined spatial mode and its wavelength can be arbitrary depending on the availability of the corresponding nonlinear processes. Although the photons are not deterministic because of the vacuum contribution, it should be applicable to a wide range of applications in quantum sensing and quantum imaging for enhancing the signal levels.

\begin{acknowledgments}
\textit{\bf Acknowledgments.} The work is supported by City University of Hong Kong (Project No.9610522), the General Research Fund from Hong Kong Research Grants Council (CityU 11307823, CityU 11215024, CityU 11204523), and Innovation Program for Quantum Science and Technology (No. 2021ZD0301500), Joint NSFC/RGC Collaborative Research Scheme (CRS-CityU103/24).

\end{acknowledgments}
\vskip 0.2in
\noindent {\bf AUTHOR DECLARATIONS}

The authors have no conflicts of interest to disclose.



%
\bibliography{bibliography}

\end{document}